# Magnetic ordering and spin dynamics in $S = 5/2$ staggered triangular lattice antiferromagnet $Ba_2MnTeO_6$

Lisi Li,[1] Narendirakumar Narayanan,[2,3] Shangjian Jin,[4] Jia Yu,[1] Zengjia Liu,[1] Hualei Sun,[4] Chin-Wei Wang,[2] Vanessa Peterson,[2] Yun Liu,[3] Danilkin Sergey,[2] Dao Xi Yao,[4] Dehong Yu,[2] and Meng Wang[1]*

[1]Center for Neutron Science and Technology, School of Physics, Sun Yat-Sen University, Guangzhou, 510275, China

[2]Australian Nuclear Science and Technology Organisation, New Illawarra Road, Lucas Heights NSW 2234, Australia

[3]Research School of Chemistry, The Australian National University, ACT 2601, Australia

[4]School of Physics, Sun Yat-Sen University, Guangzhou, 510275, China

(Dated: June 5, 2020)

We report studies of the magnetic properties of a staggered stacked triangular lattice $Ba_2MnTeO_6$ using magnetic susceptibility, specific heat, neutron powder diffraction, inelastic neutron scattering measurements, as well as first principles density functional theory calculations. Neutron diffraction measurements reveal an antiferromagnetic order with a propagated vector $\boldsymbol{k} = (0.5, 0.5, 0)$ and Néel transition temperature of $T_N \approx 20$ K. The dominant interaction derived from the Curie-Weiss fitting to the inverse DC susceptibility is antiferromagnetic. Through modelling the INS spectrum with the linear spin wave theory, the magnetic exchange interactions for the nearest intralayer, nearest interlayer, and next nearest interlayer are determined to be $J_1 = 0.27$ (3), $J_2 = 0.27$ (3), and $J_3 = -0.05$ (1) meV, respectively, and a small value of easy-axis anisotropy of $D_{zz} = -0.01$ meV is introduced. We derive a magnetic phase diagram that reveals that it is the competition between $J_1$, $J_2$, and $J_3$ that stabilizes the collinear stripe-type antiferromagnetic order.

## I. INTRODUCTION

Geometrically frustrated magnets have attached much attention due to their novel low temperature states, such as spin ice, spin liquid, and noncollinear magnetic states[1-4]. The triangular lattice is a representative geometrically frustrated structure, and although simple, exhibits a diversity of ground states[5,6]. For a two-dimensional Heisenberg triangular lattice antiferromagnet with dominant nearest neighboring intralayer coupling, a ground state with a 120° spin structure within the in-plane is realized, such as in a number of materials, such as in $Ba_8MnNb_6O_{24}$[7], $Rb_4Mn(MoO_4)_3$[8], $Ba_3MnSb_2O_9$[9], and $Ba_2La_2MTe_2O_{12}$ (M=Co, Ni)[10,11]. When anisotropy and further-neighbor magnetic coupling arise, the ordered state that finally occurs is a consequence of a subtle balance among these factors[12]. In $CuCrO_2$, an incommensurate magnetic structure is stabilized by a combination of the coupling between adjacent planes, the anisotropic in-plane nearest-neighbor interlayer interactions, and the weak antiferromagnetic (AF) next-nearest-neighbor interaction[13–15]. While for $CuFeO_2$, a three-dimensional collinear magnetic structure is stabilized by strong third-neighbor intralayer coupling and an interlayer coupling[16,17]. Strong easy axis anisotropy also plays an important role in determining the magnetic ground state, such as in 2H-$AgNiO_2$, which displays a collinear alternating stripe magnetic structure[18–20]. Therefore, the triangular lattice provides a playground for exploring exotic magnetic ground states.

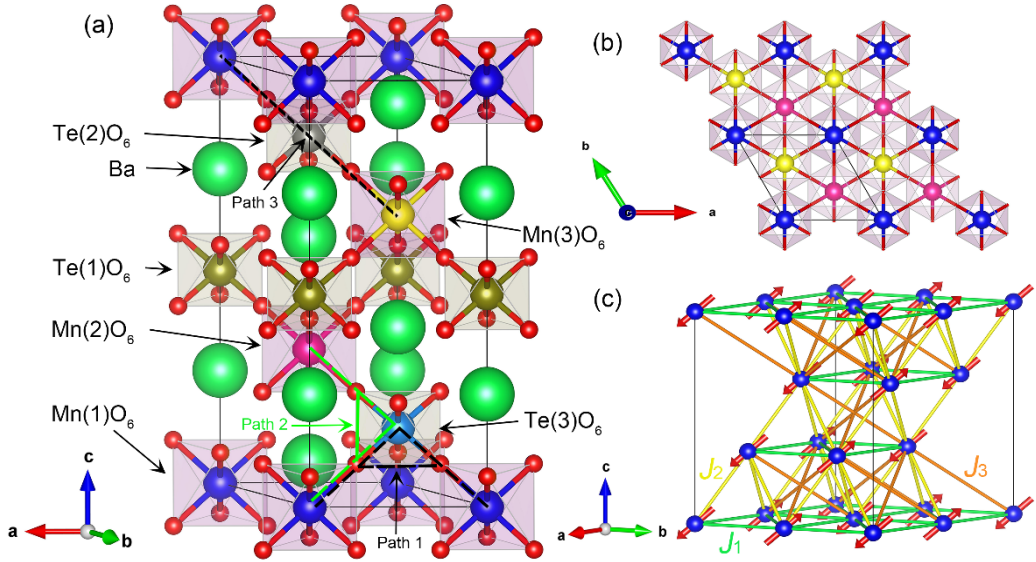

FIG. 1: (a) A unit cell of $Ba_2MnTeO_6$. The dashed curves represent for the superexchange coupling paths of $Mn^{2+}$-$O^{2-}$-$O^{2-}$-$Mn^{2+}$ and $Mn^{2+}$-$O^{2-}$-$Te^{6+}$-$O^{2-}$-$Mn^{2+}$ for the nearest intralayer (Path 1) and interlayer (Path 2), and the next-nearest interlayer (Path 3) coupling with path of $Mn^{2+}$-$O^{2-}$-$Te^{6+}$-$O^{2-}$-$Mn^{2+}$. (b) The layer structure of $MnO_6$ octahedron viewed along the $c$ axis. (c) A sketch of the magnetic structure of $Ba_2MnTeO_6$. An exchange coupling network for $J_1$, $J_2$, and $J_3$ shown as the chemical bonds of pairs of $Mn^{2+}$ ions in green, yellow, and orange, respectively.

In $Ba_2MnTeO_6$, $Mn^{2+}$ ions form a uniform triangular layer in the $ab$ plane and stack along the $c$ axis[21]. The triangular lattice shifts to the center of the triangle of the neighboring triangular lattice viewing along the $c$ axis as shown in Fig. 1(b). As a result, the triangular layers are stacked as $ABCABC\cdots$ along the $c$ axis as shown in Fig. 1(a), where $A$, $B$, and $C$ are the Mn (1), Mn (2), and Mn (3) layers, respectively. The structure is similar to $Ba_2CoTeO_6$. While for $Ba_2CoTeO_6$, the nearest interlayer distance of $Co^{2+}$ in the stacking $B$ and $C$ layers is shorter than that of the nearest distance between the $A$ and $B$ layers. Therefore, the $B$ and $C$ layers form a so-called bulked honeycomb lattice that is isolated with the $A$ layer[22–24]. In $Ba_2MnTeO_6$, the interlayer distance is the same in all the neighboring triangular layers. Moreover, the distance of the nearest intralayer Mn-Mn (5.7533(6) Å) is almost the same as the nearest interlayer distance (5.7566(6) Å). In the staggered stacked triangular lattice, the stacking geometry is characterized by the ratio of the nearest interlayer distance of neighboring layers to that of the intralayer distance, and critical to magnetic properties[25]. For example, this ratio in two-dimensional $NiGa_2S_4$ is 3.31[26–28], and in the triangular arrangement of spin chain $CsNiCl_3$ is 0.41[29–32]. The ratio in $Ba_2MnTeO_6$ is close to 1. This motivates us to explore whether the compound behaves like a two-dimensional layer system or a one-dimensional spin chain system. To our knowledge, no experimental research has been conducted to explore the magnetic properties of $Ba_2MnTeO_6$ yet. Therefore, an experimental investigation into the role of the interlayer and intralayer couplings on magnetic properties is of great importance.

In this paper, we investigate the magnetic properties of $Ba_2MnTeO_6$ by combining magnetic susceptibility, specific heat, neutron powder diffraction, and inelastic neutron scattering measurements with density functional theory calculations. We find that $Ba_2MnTeO_6$ exhibits a three-dimensional stripe-type collinear AF order with a propagation vector of $\mathbf{k} = (0.5, 0.5, 0)$ below the magnetic phase transition temperature $T_N \approx 20$ K. By modelling the INS spectrum, a ferromagnetic (FM) next-nearest interlayer coupling $J_3$ is needed to reproduce the experimental spectrum in addition to the nearest intralayer coupling $J_1$ and nearest interlayer couplings $J_2$. Moreover, a small easy axis anisotropy $D_{zz}$ is introduced which originates from the strong dipolar interaction that typically presents in large spin systems. We discuss the effect of these exchange couplings for stabilizing the magnetic ground state of $Ba_2MnTeO_6$.

## II. EXPERIMENT AND CALCULATION DETAILS

High-quality powder samples of $Ba_2MnTeO_6$ were synthesized by a conventional solid-state reaction. Stoichiometric starting materials $BaCO_3$ (99.99%), $MnCO_3$ (99%), and $TeO_2$ (99.99%) were ground thoroughly in an agate mortar and then pressed into a pellet. The pellet was calcined in 1150°C for 6 days with several intermediate grindings to obtain a highly homogeneous powder. No impurity phase was detected in X-ray diffraction (XRD) data. However, we found peaks from impurity phase $Mn_3O_4$ in neutron powder diffraction (NPD) pattern corresponding to $\sim 1wt.\%$ [33]. Zero-field-cooled (ZFC) and field-cooled (FC) Magnetic susceptibility and specific-heat measurements were performed using a physical property measurement system (PPMS, Quantum Design).

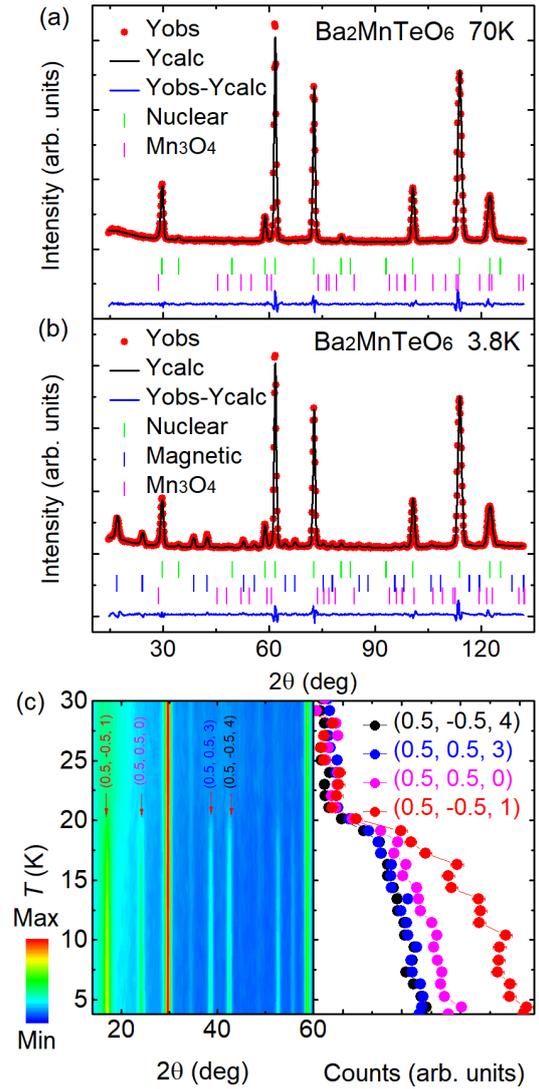

FIG. 2: Calculated and observed neutron diffraction data for $Ba_2MnTeO_6$ measured at (a) 70 K, and (b) 3.8 K. (c) Left: Temperature dependence of NPD data in the range 3.8 to 30 K measured with a step 1 K. Right: Temperature dependence of the integrated intensities for (0.5, −0.5, 1), (0.5, 0.5, 0), (0.5, 0.5, 3), and (0.5, −0.5, 4) magnetic peaks.

TABLE I: Fractional atomic coordinates, Wyckoff sites, and selected bond paths and bond angles of $Ba_2MnTeO_6$ at 70K determined by neutron powder diffraction. Refined profile factors are $R_p$ = 9.47%, $R_{wp}$ = 9.28%, $\chi^2$= 2.94. All crystallographic sites are fully occupied. The numbers in the brackets indicate errors.

| Ions | Wyckoff site | x/a | y/b | z/c |
|---|---|---|---|---|
| $Ba^{2+}$ | 3a | 0.0000 | 0.0000 | 0.2640(14) |
| $Mn^{2+}$ | 6c | 0.0000 | 0.0000 | 0.0000 |
| $Te^{4+}$ | 3b | 0.0000 | 0.0000 | 0.5000 |
| $O^{2-}$ | 18h | -0.1758(8) | -0.3516(16) | -0.0906(9) |

| Bond path | Lengths (Å) | Bond angle | Angles(deg) |
|---|---|---|---|
| Mn(1)-Mn(1) ($J_1$) | 5.7533(6) | Mn(1)-O-Te(3) | 178.1(7) |
| Mn(1)-Mn(2) ($J_2$) | 5.7566(6) | O(4)-Te(3)-O(6) (path 1) | 91.4(5) |
| Mn(1)-Mn(2) ($J_3$) | 8.1387(7) | O(4)-Te(3)-O(17) (path 2) | 88.6(5) |

NPD experiments were conducted on the high intensity powder diffractometer Wombat installed at the OPAL reactor, ANSTO[34]. Data were collected from 3.8 to 70 K with a neutron wavelength of $\lambda$ = 2.4124 Å. Powder samples were loaded into a cylindrical vanadium can with an aluminum cap. NPD data were analyzed by employing the Rietveld method using the *FullProf* Suite software[35, 36].

Inelastic neutron scattering (INS) experiments were performed on the cold-neutron time-of-fight spectrometer Pelican at the OPAL reactor, ANSTO[37]. Incident neutron energies were $E_i$ = 3.7 and 14.8 meV with the corresponding energy resolutions of △E = 0.14 and 0.35 meV, that is determined by the full width at half maximum of the elastic peak from a vanadium standard sample. Powder samples were loaded into an annular aluminum can and measured at 1.5 K. The background from an empty aluminum can was measured and subtracted. The data analysis was performed using the program LAMP[38]. Linear spin wave theory was employed to model the INS spectra using the *SpinW* software[39].

First-principles density functional theory (DFT) calculations were performed using full-potential linearized augmented plane-wave plus the local orbital method as implemented in the WIEN2K code[40]. The value of $R_{MT}K_{max}$ was set to 6. The mesh of special $k$ points is selected to be 2 × 2 × 2. The selected unit cell is a supercell (magnetic unit cell) containing 120 ions including 12 independent $Mn^{2+}$ ions. The crystal structure parameters and AF structure were refined against the NPD data. The density of state was calculated using the local spin density state (LSDA)[41] and LSDA+$U$ (FLL)+$J$[42]. A Coulomb repulsion $U$ was added in the fully localized limit (FLL) and $J$ considering the exchange hole contribution. The magnetic dipole-dipole interaction (MDDI) tensor was calculated using the code McPhase[43]. For the 12 $Mn^{2+}$ ions in the magnetic unit cell, contributions from Mn neighbours up to 20 Å were considered. The MDDI energies for different orientations of magnetic moments were calculated by solving an eigenvalue problem as shown in Ref [44].

## III. CRYSTAL AND MAGNETIC STRUCTURE

The crystal structure of $Ba_2MnTeO_6$ refined from single crystal X-ray diffraction data could be described by hexagonal symmetry with a layered triangular lattice in the space group $R\bar{3}m$ (No. 166)[21]. We used this as a starting structure to refine against our 70 K NPD data. An impurity phase of $Mn_3O_4$ was considered in the refinement profile in Fig. 2(a), and the result shows that this is present at about $1wt.\%$. The lattice parameters are determined to be $a = b = 5.7534$ (4) Å, and $c = 14.1052$ (2) Å, $\alpha = \beta = 90°$, and $\gamma = 120°$. Details of the atom coordinates and some selected bond distances and bond angles are shown in Table I.

The magnetic order of the material was investigated using NPD at 3.8 K. Clear magnetic peaks appear in the neutron diffraction data in Fig. 2(b). A plot of the magnetic peak intensities against temperature reveals the Néel temperature $T_N \approx 20$ K in Fig. 2(c). To determine the magnetic structure, we used a hexagonal crystal unit cell to index peaks in the NPD data at 3.8 K. All magnetic peaks could be indexed with a propagation vector of $\boldsymbol{k} = (0.5, 0.5, 0)$. Magnetic peaks at $2\theta = 17.0°$, $24.1°$, $38.6°$ and $42.5°$ correspond to $(H, K, L) = (0.5, -0.5, 1)$, $(0.5, 0.5, 0)$, $(0.5, 0.5, 3)$, and $(0.5, -0.5, 4)$, respectively. Here, the momentum transfer $|Q|$ is calculated from Miller indices (H, K, L) using the relation $|Q| = 2\pi \sqrt{\frac{4(H^2+HK+K^2)}{3a^2} + \frac{L^2}{c^2}}$ in the $R\bar{3}m$ space group where $a$ and $c$ are the lattice parameters.

TABLE II: The irreducible representations and corresponding basis vectors for the space group $R\bar{3}m$ with $\boldsymbol{k} = (0.5, 0.5, 0)$ and Mn coordination (0,0,0), and the corresponding refined magnetic profile factors $R_{mag}$.

| IR | BV | Mn | $R_{mag}$ |
|---|---|---|---|
| $\Gamma_1^1$ | $\psi_1$ | [1, 1, 0] | 6.54% |
| $\Gamma_1^2$ | $\psi_2$ | [1, -1, 0] | 5.49% |
|  | $\psi_3$ | [0, 0, 1] |  |

Furthermore, we performed Rietveld refinement of the structure based on representational analysis using the program BasIreps within the FullProf package[35,45,46]. The representational analysis for the propagation vector $\boldsymbol{k} = (0.5, 0.5, 0)$ and the space group $R\bar{3}m$ gives two nonzero irreducible representations (IR) for one magnetic site of Mn (0,0,0).

$$\Gamma_{mag}^{Mn} = \Gamma_1^1 + \Gamma_1^2 \quad (1)$$

The basis vector of the two IRs of Mn is displayed in Table II. We assume that the magnetic structure can be described by a single IR, one of two basic vectors: (i): $\psi_1$ and (ii): the linear combination of $\psi_2$ and $\psi_3$. The magnetic structures represented by the basic vectors (i) and (ii) are assigned to model (i) and model (ii), respectively. We performed Rietveld refinement of these two models against our data, revealing model(ii) having the better agreement of $R_{mag} = 5.49\%$ with a moment direction tilted from the $ab$ plane at an angle of about 41°. From the NPD experiment, we could not conclusively determine the magnetic structure, given the fact that the two magnetic $R$ factors are close. We note that a magnetic structure with $c$ component is more favorable in a layer structure with considerable interlayer coupling, such as in $CuCrO_2$[14].

Figure 2(b) displays the observed and calculated NPD data at 3.8 K using the model (ii) with profile factors of $R_p$ = 10.7%, $R_{wp}$ = 10.4%, $R_e$ = 5.43%, and $\chi^2$ = 3.67 along with an impurity phase Mn$_3$O$_4$. The magnetic moment of Mn$^{2+}$ with a $c$ axis component and tilting at an angle of 41° from the $ab$ plane is determined to be 4.49 $\mu_B$, which is close to the full moment of 5 $\mu_B$ in the high spin state of $S = 5/2$. The reduction of moment could be attributed to the bonding effect and geometrical frustration as shown in Fig. 1(c).

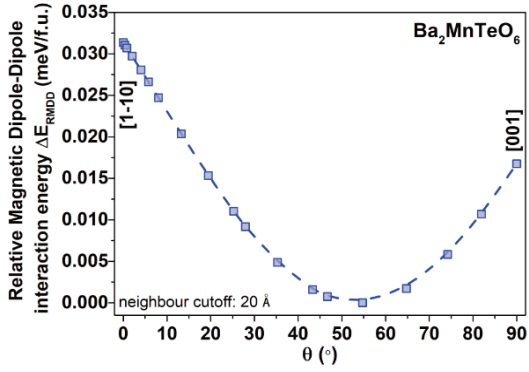

FIG. 3: Relative MDDI energy for different angles of the Mn$^{2+}$ magnetic moments tilting from [1, −1, 0] towards the $c$ axis. The energy minimum is around 55°.

Investigation of the magnetic anisotropy may lead to resolving the magnetic structure. Magnetic anisotropy can be determined using the MDDI for systems with a large spin and weak spin-orbit coupling (SOC)[44,47]. Ideally, for Mn$^{2+}$ in high spin state $S = 5/2$, the SOC should vanish and the MDDI is maximum in open $d$ electron in Ba$_2$MnTeO$_6$[48]. Therefore, we model the MDDI energy as described in the experimental section for a tilting ($\theta$) of the magnetic moments from the [1, −1, 0] direction on the $ab$-plane towards the $c$ axis. The relative MDDI energy $\Delta E_{RMDD}$ exhibits a minimum around 55° from the $ab$-plane which is in good agreement with the experimental result of 41° as shown in Fig. 3. Therefore, we determine the magnetic structure to be collinear stripe-type AF order corresponding to the model (ii). We note that neutron diffraction measurements using a single crystals is needed for confirm the magnetic structure experimentally.

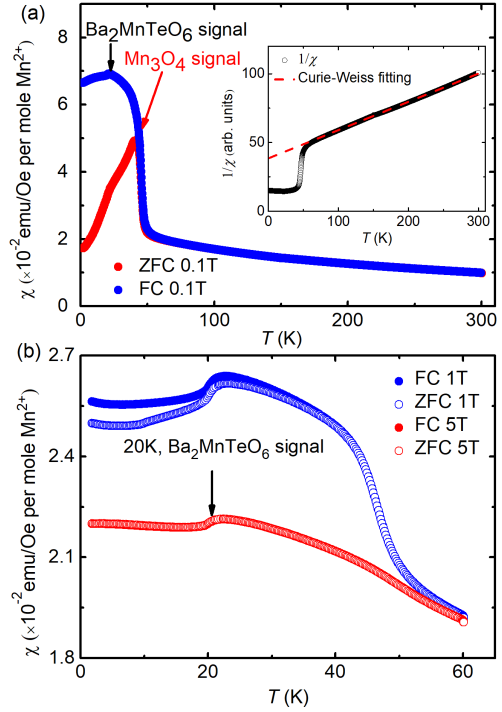

FIG. 4: (a) Field cooling (FC) and zero field cooling (ZFC) magnetic susceptibility measured at $\mu_0 H$ = 0.1 T. The inset shows the inverse susceptibility and a high temperature Curie-Weiss fitting. (b) FC and ZFC magnetic susceptibility measured at $\mu_0 H$ = 1 T and 5 T.

## IV. MAGNETIC SUSCEPTIBILITY

Figure 4(a) shows temperature dependence of ZFC and FC DC susceptibility $\chi$ ($T$) data of Ba$_2$MnTeO$_6$ measured with a magnetic field of $\mu_0 H$ = 0.1 T. Divergence of the ZFC and FC susceptibility at ~41 K could be attributed to the impurity phase Mn$_3$O$_4$ which orders ferrimagnetically below this temperature[49,50]. The susceptibility measured at low magnetic field is sensitive to the ferrimagnetic signal of the small amount of Mn$_3$O$_4$ impurity, which has been reported in other $Mn^{2+}$ contained compounds such as $Mn_2OBO_3$[51], $Sr_2MnTeO_6$[52], and $Mn_4Ta_2O_9$[47]. At around 20 K, a small kink appears in the ZFC and FC data which arises from the intrinsic AF transition of $Ba_2MnTeO_6$, consistent with the NPD data. For the higher magnetic fields of $\mu_0H$ = 1 and 5 T as shown in Fig. 4(b), the divergence at ~41 K is weakened while the intrinsic magnetic transition at 20 K could be more clearly recognized.

The magnetic susceptibility $\chi$ ($T$) above $T_N$ ≈ 20 K agrees with the Curie-Weiss law $\chi = \chi_0 + C/(T - \Theta_{CW})$, where $\chi_0$ is a contribution from diamagnetism and Van Vleck paramagnetism, $C$ is the Curie constant, and $\Theta_{CW}$ is the Curie-Weiss temperature[53]. Fitting of data between 75 and 300 K yields $\chi_0$= −6.5(5) ×$10^{-4}$ emu $Oe^{-1}mol^{-1}$, $C$ = 4.38(3) emu K $mol^{-1}$, and $\Theta_{CW}$ = −168(2) K. The negative Curie-Weiss temperature indicates that the dominant exchange interaction of $Ba_2MnTeO_6$ is antiferromagnetic. The effective moment is estimated to be $\mu_{eff}$= 5.88(1) $\mu_B$, close to the theoretical effective moment $\mu_{eff}=g\sqrt{S(S+1)}$ = 5.91 $\mu_B$ for the high spin configuration of $Mn^{2+}$ with $S = 5/2$. The frustrated parameter obtained from the estimation of an empirical formula $f = |\Theta_{CW}|/T_N$ is 8.43 in $Ba_2MnTeO_6$, which is lower than that of a strong frustrated system with $f \geq 10$ [4].

## V. SPECIFIC HEAT

Specific heat measurements against temperature are presented in Fig. 5(a). A sharp λ-like transition occurs at the magnetic transition temperature ~20 K. The magnetic contribution to the specific heat was extracted by subtracting the phonon contribution from the total specific heat. For this purpose, a modified Debye model considering the existence of two phonon spectra was employed to fit the high temperature data from 55 to 200 K. The phonon contribution is extrapolated down to 1.8 K[52]. The modified Debye model follows the formula[54]:

$$C_{ph} = 9R \sum_{n=1}^{2} C_n \left(\frac{T}{\Theta_{Dn}}\right)^3 \int_0^{\frac{\Theta_{Dn}}{T}} \frac{x^4 e^x}{(e^x - 1)^2} dx, \quad (2)$$

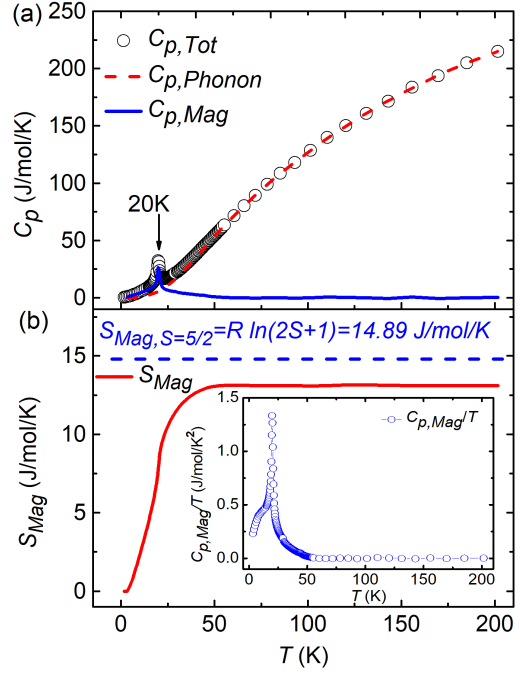

FIG. 5: (a) Specific heat $C_p$ measured at zero magnetic field. The red dashed line is the phonon contribution fitted using a modified Debye model. The blue solid line represents the magnetic contribution. (b) The red solid line is the magnetic entropy $S_{mag}$ and the blue dashed line marks the expected magnetic entropy for $Mn^{2+}$ with $S= 5/2$. The inset shows the magnetic contribution of the heat capacity divided by $T$, $C_{p, mag}/T$.

The fitting results indicate that, of 10 atoms in the formula unit, 4.5 atoms have a Debye temperature $\Theta_{D1}$ of 249 ± 4 K and 5.5 atoms have a Debye temperature $\Theta_{D2}$ of 753±20 K, close to the ratio at 4 : 6 of the heavy atoms ($Ba^{2+}$, $Mn^{2+}$, $Te^{6+}$) to the light atoms ($O^{2−}$) for $Ba_2MnTeO_6$.

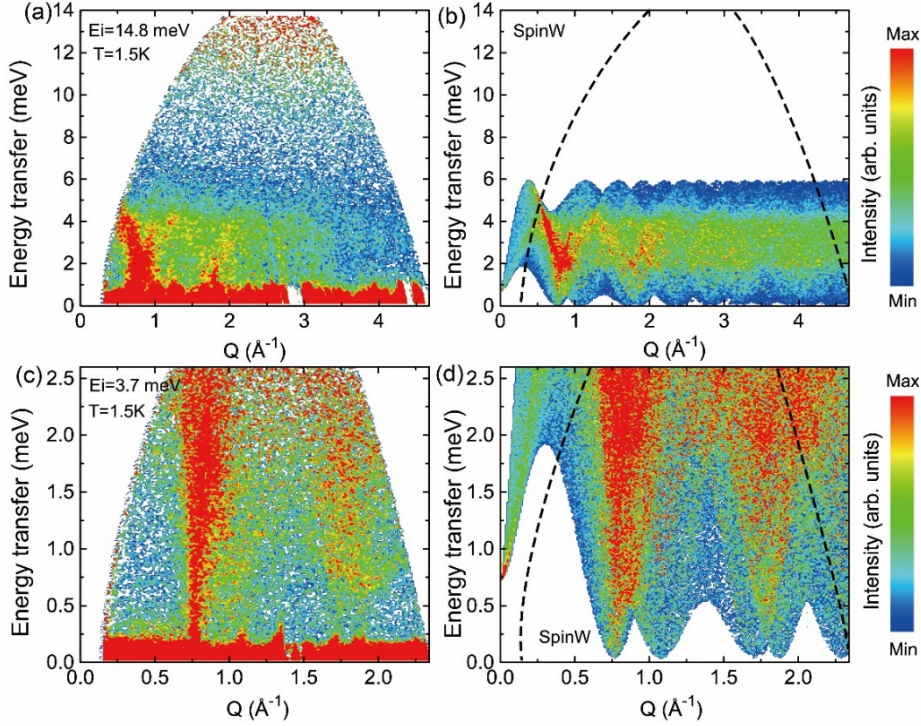

FIG. 6: INS spectra measured with incoming neutron energies of (a) $E_i$ = 14.8 meV and (c) $E_i$ = 3.7 meV at $T$ = 1.5 K. (b) and (d) Simulated powder-averaged INS spectra using *SpinW* with $J_1$ = 0.27, $J_2$ = 0.27, $J_3$ = −0.05, and $D_{zz}$ = −0.01 meV. The instrument resolutions (b) $\triangle E$ = 0.35 meV and (d) $\triangle E$ = 0.14 meV are convoluted. Dashed lines in (b) and (d) represent the $Q − E$ space of the instrument.

The magnetic entropy $S_{mag} = \int C_{p,mag}/T\, dT$ is attributed to the magnetic state change, which yields $S_{mag}$ = 13.11 J/mol/K at 200 K as shown in Fig. 5(b). The value is close to the theoretical expected value for $S = 5/2$, where $S_{mag}$ = R·ln (2$S$ + 1) = 14.89 J/mol/K. The reduction of $S_{mag}$ could arise from the magnetic frustration and bonding effects as consistence with the reduced ordered moment refined from the NPD data.

## VI.   MAGNETIC EXCITATIONS

Figures 6(a) and (c) show the INS spectra collected at 1.5 K with the incident energies of $E_i$ = 14.8 meV and $E_i$ = 3.7 meV, respectively. The excitations exhibit clear dispersions and the intensities decrease with increasing |$Q$|, demonstrating a magnetic scattering origin. The momentums of the magnetic excitations at |$Q$| = 0.77, 1.09, 1.72, and 1.88 Å$^{-1}$ correspond to the magnetic Bragg peaks of (H, K, L) =(0.5, −0.5, 1), (0.5, 0.5, 0), (0.5, 0.5, 3) and (0.5, −0.5, 4), respectively. No spin gap is observed in either spectra, possibly because of instrumentation limitations or its gapless nature. For a further understanding of the powder averaged spectra, we turn to linear spin wave theory using the *SpinW* package.

During the simulation, we take into account the exchange couplings $J_1$, $J_2$, $J_3$, and an easy-axis anisotropy $D_{zz}$, which construct the typical Hamiltonian:

$$\hat{H} = \sum_{i,j} J_{i,j} \mathbf{S}_i \cdot \mathbf{S}_j + D_{zz} \sum_{i,z} S_{i,z}^2, \quad (3)$$

where $J_{i,j}$ denotes the exchange coupling parameters, as shown in Fig. 1(c). The second term is the easy-axis anisotropy,

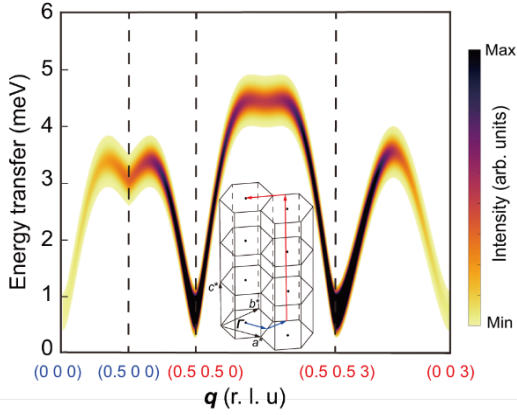

FIG. 7: Simulated spin waves along high symmetry directions illustrated as the blue and red paths in the 3D Brillouin zone as shown in the inset. Intensities are represented by the color.

which would lead to a spin gap. Note that $S_{i,z}$ is the component of spin operator in the rotating frame, whose $z$-axis is along the direction of classical spin $S_i$.

A significant contribution from the MDDI is expected to be a driving force in determining the direction of the magnetic moments in $Ba_2MnTeO_6$, and we employ a small value of $D_{zz} = -0.01$ meV as the easy-axis anisotropy term as $Mn_4Ta_2O_9$ in simulation[47]. The nearest intralayer coupling $J_1$ with the Mn-Mn distance 5.7533(6) Å and the nearest interlayer coupling $J_2$ with the Mn-Mn distance 5.7566(6) Å occur through $Mn^{2+}$-$O^{2-}$-$O^{2-}$-$Mn^{2+}$ and $Mn^{2+}$-$O^{2-}$-$Te^{6+}$-$O^{2-}$-$Mn^{2+}$ superexchange paths as drawn in Fig. 1(a). According to the Goodenough-Kanamori rule, the coupling through the former path is antiferromagnetic[55]. For the latter, it is a nearly 90° path in Table I, which should be also antiferromagnetic because the filled outermost orbital is the $4d$ orbital of $Te^{6+}$ ions according to other compounds with similar path[10,11,56–58]. Therefore, both $J_1$ and $J_2$ should be AF with comparable coupling strengths. The next-nearest interlayer coupling $J_3$ is also considered. The $J_3$ corresponds to the superexchange path of $Mn^{2+}$-$O^{2-}$-$Te^{6+}$-$O^{2-}$-$Mn^{2+}$ and the Mn-Mn distance of 8.1387(7) Å as illustrated in Fig. 1(a). The $J_3$ could be either antiferromagnetic as in $Sr_2CuTeO_6$[57] or ferromagnetic as in $Sr_2MnTeO_6$[52].

By comparing against experimental data, we determine the strengths of the exchange couplings $J_1$=0.27 (3), $J_2$=0.27 (3), $J_3$=−0.05(1), and $D_{zz}$= −0.01 meV. The reproduced spectra are presented in Figs. 6(b) and (d), respectively. The powder INS spectra can be well described by the set of exchange couplings. A spin gap around $E$ = 0.05 meV emerges that is beyond the instrument resolution. We plot the dispersion relations based on the couplings along the high symmetry directions in the 3D Brillouin zone, as displayed in Fig 7. This plot would be useful for a comparison with the spin waves measured from single crystal samples.

## VII.    MAGNETIC PHASE DIAGRAM

To quantify the effect of these exchange couplings in stabilizing the magnetic ground state of $Ba_2MnTeO_6$ and materials with the similar structures, a magnetic phase diagram is derived by considering the exchange couplings $J_1$, $J_2$ and $J_3$ based on a pure Heisenberg model without considering the $D_{zz}$. We assume the $J_1 > 0$, $J_2 > 0$, and the sign of $J_3$ is variable. Moreover, we define $\alpha = J_2/J_1$, $\beta = J_3/J_1$ for visualization. Here, we consider a magnetic ground state for the system with a single propagation vector $q$ defined as:

$$q = q_a a^* + q_b b^* + q_c c^*, \quad (4)$$

where $a^*$, $b^*$, and $c^*$ are the reciprocal space basis vectors corresponding to the basis vectors $a$, $b$, and $c$ in real space as shown in Figs. 8(b) and (c). The coordinate corresponds to the crystal structure of $Ba_2MnTeO_6$ in Fig. 1(a).

The ground state for a classical spin system only contains component from $S_z$, thus the product of $S_i$ and $S_j$ is given by:

$$S_i \cdot S_j = S^2 \cos q \cdot (r_i - r_j), \quad (5)$$

where $S$ represents the value of the spin vector $S$.

By combining Eqs. 3, 4, and 5, the ground state energy $E_0$ is obtained as a function of $q_a$, $q_b$, $q_c$, $\alpha$, and $\beta$:

$$\frac{E_0}{3J_1 N S^2} =$$
$$2\cos(\pi(q_a + q_b))\cos(\pi(q_b - q_a))$$
$$+\cos(2\pi(q_a + q_b))$$
$$+ 2\alpha \cos(\pi(q_a + q_b)) \cos(2\pi((q_b - q_a)/6 - q_c/3))$$
$$+\alpha \cos 2\pi((q_b - q_a)/2 + q_c/3)$$
$$+ \beta \cos 2\pi(2(q_b - q_a)/3 - q_c/3)$$
$$+2\beta \cos 2\pi(q_a + q_b) \cos 2\pi((q_b - q_a)/3 - q_c/3). \quad (6)$$

In Fig. 8(a), numerical analysis is employed for the $q$ values within the first Brillouin zone and $0 \leq \alpha \leq 2$ and $-1 \leq \beta \leq 1$ to determine the ground state with the minimum $E_0$[59–61]. Four phases are identified as presented. Stripe, helical 1, helical 2 phases correspond to the propagation vectors of $q = (0.5, 0.5, 0)$, $(-0.5, 0.5, 1/2)$, and $(0, 1, 1/2)$, respectively. The spiral phase exhibits a noncollinear incommensurate order. When both interlayer couplings $J_2$ and $J_3$ are weak ($\alpha \approx \beta \approx 0$), the magnetic structure stabilizes with a 120° angle between the moments, corresponding to the propagation vector $q'_3 = (1/3, 1/3, 0)$. The FM $J_3$ is crucial for realization of the collinear stripe-type AF order. Ba$_2$MnTeO$_6$ with the values of $\alpha = J_2/J_1 \approx 1.0$, $\beta = J_3/J_1 \approx -0.18$ is within the stripe phase as marked in Fig. 8(a). Many two-dimensional magnetic materials such as Ba$_8$MnNb$_6$O$_{24}$[7], Rb$_4$Mn(MoO$_4$)$_3$[8], Ba$_3$MnSb$_2$O$_9$[9], and Ba$_2$La$_2$MTe$_2$O$_{12}$ (M=Co, Ni)[10,11] with weak interlayer couplings ($\alpha \approx \beta \approx 0$) show a coplanar 120° magnetic order in the spiral phase. For the helical 1 phase where $J_3$ is antiferromagnetic and $0 < J_2 \leq J_1$, the in-plane AF coupling $J_1$ dominants and results in AF in-plane order. However, the intralayer AF coupling $J_2$ is strong in the helical 2 phase, yielding FM in-plane order.

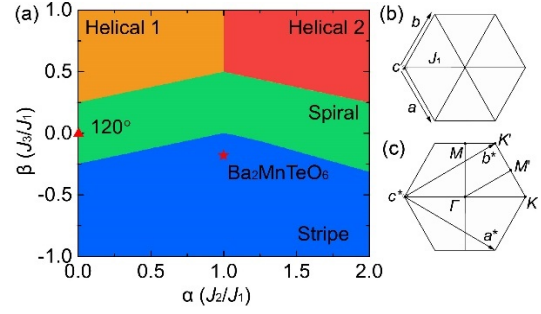

FIG. 8: (a) A magnetic phase diagram with variable exchange couplings for structures similar to Ba$_2$MnTeO$_6$. Helical 1 and helical 2 represent the magnetic structures with propagation vectors $q_1 = (-0.5, 0.5, 1/2)$ and $q_2 = (0, 1, 1/2)$, respectively. The term spiral phase refers to an incommensurate magnetic order. The stripe phase has the propagation vector $q_3 = (0.5, 0.5, 0)$ with a stripe-type AF order. The ★ in the stripe phase shows the position of Ba$_2$MnTeO$_6$, and the filled △ indicates triangular layered materials with weak interlayer couplings. (b) Sketch of the triangular lattice. The lattice basis vectors are donated by $a$, $b$, and $c$. $J_1$ refers to the nearest-intralayer coupling. (c) The first Brillouin zone for the triangular lattice. $a^*$, $b^*$, and $c^*$ refer to the basis vectors.

## VIII. DFT CALCULATION

Figure 9 show the total DOS and partial DOS of Mn 3$d$ bands calculated using the LSDA and LSDA+$U$(FLL)+($J$) methods, respectively. The complete occupation of the majority spin is consistent with the 3$d^5$ electron configuration in the high spin state $S = 5/2$ of Mn$^{2+}$. Bonding to anti-bonding splitting is visible as mask by the arrows in Figs.9 (b) and (c). Anti-bonding bands are further split into the $e_g$ ($d_{z^2}$, $d_{x^2-y^2}$) and $t_{2g}$ ($d_{xy}$, $d_{xz}$, $d_{yz}$) manifolds owing to crystal splitting which is visible clearly for LSDA, but no gap between them due to the strong Hund's coupling. Upon the adding of an orbital-dependent correction to the on-site Coulomb repulsion using the LSDA+$U$(FLL)+($J$) method, where we only consider the fully localized limit with typical values for $U$ (3.81 eV=0.28 Ry) and $J$ (0.75 eV=0.055 Ry) taken from ref[62], band gap increases and the $d$-band states become smeared. Here, $J$ accounts for the already considered repulsion between the parallel spins due to the Pauli principle. Adding $U$ shifts the occupied spin-up $d$ bands (both $e_g$ and $t_{2g}$) by $-(U-J)/2$ and the unoccupied spin-down $d$ bands by $+(U-J)/2$.

Calculated spin moments using LSDA and LSDA+$U$(FLL)+$J$ are 4.27$\mu_B$/Mn$^{2+}$ and 4.43$\mu_B$/Mn$^{2+}$, respectively. The difference to 5 $\mu_B$/Mn$^{2+}$ comes from the bonding effects, mainly attributed to hybridization with the O 2$p$ orbitals. The order magnetic moment size calculated with LSDA+$U$(FLL)+($J$) is in good agreement with the NPD result of 4.49 $\mu_B$/Mn$^{2+}$ indicating a strong electron-electron repulsion correction in the system and a very small orbital moment. The latter is an indicator of very weak spin-orbit coupling in this compound and that the magnetic anisotropy is dominated by the MDDI. For the total DOS as shown in Fig 9, the band gap using LSDA increases from 1.05 eV to 1.85 eV upon using LSDA+$U$(FLL)+($J$). The increase is due to the electron-electron repulsion correction which demonstrates that Ba$_2$MnTeO$_6$ could be classified as a Mott insulator.

## IX. SUMMARY

In summary, we carried out susceptibility, specific heat, and neutron scattering experiments to investigate the magnetic properties of the staggered stacked triangular lattice Ba$_2$MnTeO$_6$. A stripe-type AF order with a Néel temperature $T_N \approx 20$ K and the propagation vector $\boldsymbol{k} = (0.5, 0.5, 0)$ is revealed. The spin waves excitations of the stripe AF order with energy transfer extending to 5 meV are observed at 1.5 K. Through modelling of the spin wave excitations based on the linear spin wave theory, we determined the magnetic interactions $J_1$= 0.27 (3), $J_2$= 0.27 (3), and $J_3$= −0.05 (1) meV and an easy-axis anisotropy term $D_{zz}$= −0.01 meV. The resultant couplings exist within the stripe phase in the magnetic phase diagram for the triangular lattice that we derived. A FM interaction of $J_3$ is the key to stabilizing the collinear stripe-type AF order.

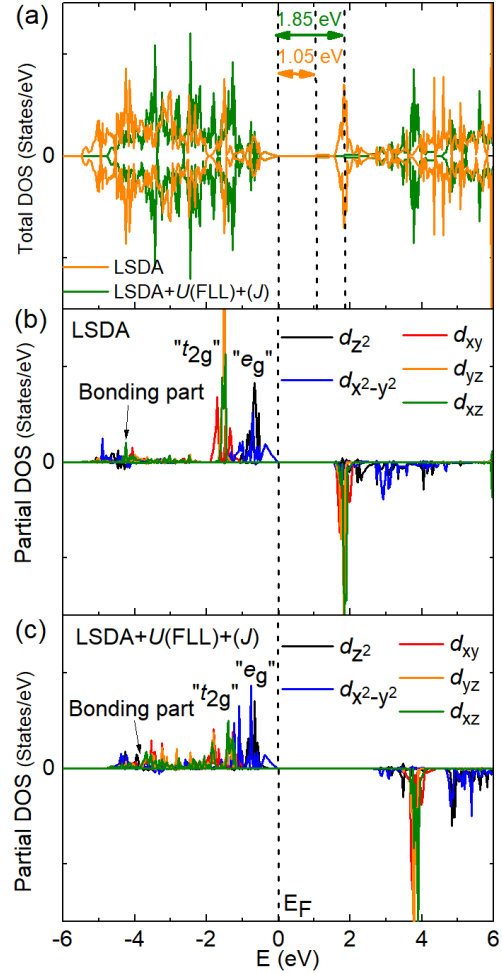

FIG. 9: (a) Total density of states (DOS) of Ba$_2$MnTeO$_6$. Partial $d$-bands DOS of Mn$^{2+}$ calculated with (b) LSDA and (c) LSDA+$U$(FLL)+($J$). $e_g$ and $t_{2g}$ represent the sum of ($d_{z^2}$, $d_{x^2-y^2}$) and ($d_{xy}$, $d_{xz}$, $d_{yz}$) bands, respectively.

## ACKNOWLEDGMENTS

The work is supported by the National Natural Science Foundation of China (Grant No. 11904414, 11904416), Natural Science Foundation of Guangdong (No. 2018A030313055), National Key Research and Development Program of China (No. 2019YFA0705700), the Fundamental Research Funds for the Central Universities (No. 18lgpy73), the Hundreds of Talents program of Sun Yat-Sen University, and Young Zhujiang Scholar program. S. J. and D. X. Y. are supported by NKRDPC-2018YFA0306001, NKRDPC-2017YFA0206203, NSFC-11974432, GBABRF-2019A1515011337, and Leading Talent Program of Guangdong Special Projects.